\documentclass[12pt,preprint]{aastex}

\slugcomment{}

\shorttitle{Study on X-ray spectra of obscured AGNs}
\shortauthors{S. Ikeda et al.}

\begin{document}


\title{Study on X-ray Spectra of Obscured AGNs based on Monte Carlo simulation - an interpretation of observed wide-band spectra 
%%Study on the connection between X-ray spectrum and structure of AGN based on Monte Carlo simulation
}


\author{Shinya Ikeda\altaffilmark{1}, Hisamitu Awaki\altaffilmark{1}, and Yuichi Terashima\altaffilmark{1}}
\affil{Department of Physics, Ehime University, Matsuyama, 790-8577, Japan}

\email{awaki@astro.phys.sci.ehime-u.ac.jp}

%\and

%\author{anyone\altaffilmark{5}}
%\affil{There or thereabout}


%% \altaffiltext{1}{Department of Physics, Ehime University, Matsuyama, 790-8577, Japan}

%\altaffiltext{1}{Visiting Astronomer, Cerro Tololo Inter-American Observatory.
%CTIO is operated by AURA, Inc.\ under contract to the National Science
%\altaffiltext{2}{Society of Fellows, Harvard University.}
%\altaffiltext{3}{present address: Center for Astrophysics,
%    60 Garden Street, Cambridge, MA 02138}
%\altaffiltext{4}{Visiting Programmer, Space Telescope Science Institute}
%\altaffiltext{5}{Patron, Alonso's Bar and Grill}


\pagebreak

\begin{abstract}
Monte Carlo simulation is one of the best tools to study the complex spectra of Compton-thick AGNs 
and to figure out the relation between their nuclear structures and X-ray spectra. 
We have simulated X-ray spectra of Compton-thick AGNs obscured by an accretion torus
whose structure is characterized by a half-opening angle, an inclination angle of the torus relative to the observer, and a column 
density along the equatorial plane.   We divided the simulated spectra into three components: 
one direct component, an absorbed reflection component and an unabsorbed reflection component. We
then deduced the dependencies of these components on the parameters describing the structure 
of the torus. Our simulation results were
applied to fit the  wide-band spectrum of the Seyfert 2 galaxy 
Mrk 3 obtained by $Suzaku$. The spectral analysis indicates that we observe the nucleus along a line of sight
intercepting the torus near its edge, 
and the column density along the equatorial plane was estimated to be 
$\sim$10$^{24}$ cm$^{-2}$.  Using this model, we can estimate the luminosities of 
both the direct emission and the emission irradiating the surrounding matter.
This is useful to find the time variability and time lag between the direct and reflected light.

\end{abstract}




%% Note that for sources with brackets in their names, e.g., [WEG2004] 14h-090,

\section{Introduction}

Active galactic nuclei (AGN) emit huge amounts of energy as
a result of accretion of matter onto 
supermassive black holes (hereafter SMBHs). Strong hard X-ray emission from these nuclei 
%%is thought to be non-thermal, and 
can be represented by power-law emission with a canonical photon 
index of ${\sim}$1.9. The strong emission irradiates material around the SMBH, leading to 
reprocessing of the emission. Observation of the reprocessed emission is conducive to revealing 
the environment of the black hole, and hence 
help to understand the problems of the fuel supply and evolution
of SMBHs.

Compton-thick AGNs are suitable for the study of the reprocessed emission in AGNs, since 
their reprocessed emission dominates over the direct emission below 10 keV due to large 
column density $N_{\rm H}>10^{24}$ $\rm{cm^{-2}}$.  In addition,
Compton-thick AGNs are thought to be abundant in the local universe \citep[e.g.,][]{risa99}. 
Thus, these objects are important for understanding some key problems in AGN research, for example synthesis 
modeling of the Cosmic X-ray Background \citep[e.g.,][]{ueda03} and the growth of AGNs.  
However, their spectra are so complex that the detailed nature of Compton-thick AGNs has thus far been unclear.  
There is another problem for spectral analyses of Compton-thick AGNs.  The baseline model 
based on previous observations, which consists of both a direct  and a reflection 
component, has been used to reproduce the complex X-ray spectra of Compton-thick AGNs 
\citep[e.g.,][]{mz95, cappi99}.  This model worked well for reproducing their spectra,
and has contributed to our understanding of AGN.
However, it is difficult to obtain information about the structure of the surrounding material from spectral 
fitting with this baseline model because the reflection model was developed for an accretion disc geometry. 
Thus, the reflection model does not match exactly with the actual reflection from the material around the black hole. 
Therefore we need a model that represents actual X-ray emission from the surrounding material using detailed Monte Carlo simulations.

Previous work that has investigated AGN X-ray spectra
by means of Monte Carlo simulations so far includes the following.  
\citet{gf91} assumed a semi-infinite, plane-parallel configuration, and they calculated reflection from a plane-parallel slab. \citet{wf99} computed transmission through a homogeneous sphere of cold material with a radius determined by its column density. 
\citet{awaki91} and \citet{ghi94} computed X-ray spectra emitted from AGNs and assumed a torus 
structure which has an open reflecting area characterized by 
its half-opening angle.
However, these results were not applied 
to make available convenient spectral-fitting models to reproduce the X-ray spectrum of,
for example, a Seyfert 2 galaxy.

We assume surrounding material with a 3D torus configuration and then simulate a spectrum 
from an AGN, considering the effect of Compton down-scattering and absorption. 
Performing Monte Carlo simulations, we investigate the relation between continuum emissions and the structure of the torus. 
Furthermore, the strength of iron line is important to reveal the existence of a hidden nucleus \citep[e.g.,][]{maiolino98}. Thus, we also investigate a dependency of the iron line 
on the structure of the torus.  We apply our simulation results to the spectral
fitting of the X-ray spectrum of Mrk 3 observed with $Suzaku$,  thereby constraining the structure of 
the torus in this Seyfert 2 galaxy.

%%%%%%%%%%%%%%%%%%%%%%%%%%%%%%%%%%%%%%%%%

\section{Model Definition and Calculation}

%% In a manner similar to \objectname authors can provide links to dataset
%% hosted at participating data centers via the \dataset{} command.  The
%% second curly bracket argument is printed in the text while the first
%% parentheses argument serves as the valid data set identifier.  Large
%% lists of data set are best provided in a table (see Table 3 for an example).
%% Valid data set identifiers should be obtained from the data center that
%% is currently hosting the data.
%%
%% Note that AASTeX interprets everything between the curly braces in the 
%% macro as regular text, so any special characters, e.g., "#" or "_," must be 
%% preceded by a backslash. Otherwise, you will get a LaTeX error when you 
%% compile your manuscript.  Special characters do not 
%% need to be escaped in the optional, square-bracket argument.

\subsection{Basic Assumptions}
In our Monte Carlo simulation, we adopted a standard spherical coordinate system with the radial 
distance ($r$) from the origin, the zenith angle (${\theta}$) from the Z-axis, and the azimuthal angle 
(${\phi}$) from the X-axis.  A primary radiation source was placed at the origin, and illuminates the  
surrounding material.
The material was assumed to be neutral and cold (T ${\leq}$ $10^{6}$ K). 

We took into account photoelectric absorption, iron fluorescence, and Compton scattering in our simulation.
%%The photoelectric absorption cross-section, ${\sigma}_{\rm{abs}}$, was calculated by summing  
%%photoelectric absorption cross-sections of H, He, C, N, O, Ne, Mg, Si, S, and Fe.
%%The cross-section of each element was deduced by using the NIST 
%% XCOM database\footnote{http://physics.nist.gov/PhysRefData/Xcom/Text/XCOM.html} and
%% the cosmic elemental abundances of \citet{ag89}. Note that the cross-section by \citet{bm92} is not valid 
%% for energies above 10 keV, and that the NIST cross-section in the 1--100 keV band is nearly equal to that by \citet{vern96} which is identified as $vern$ in $XSPEC$ (see Figure 1).  
The photoelectric absorption cross-section, ${\sigma}_{\rm{abs}}$, was calculated by using the NIST
XCOM database \footnote{http://physics.nist.gov/PhysRefData/Xcom/Text/XCOM.html}
and
the cosmic elemental abundances of \citet{ag89}.
We also calculated the photoelectric absorption cross-section of Fe
for deciding iron absorption events.
Note that the cross-section by \citet{bm92} is not valid
for energies above 10 keV, and that the NIST cross-section in the 1--100
keV band is nearly equal to that by
\citet{vern96} which is identified as $vern$ in $XSPEC$ (see Figure 1).
The Compton scattering cross-section, ${\sigma}_{\rm{es}}$, was calculated with the Klein-Nishina formula. 
The number density of electrons, $n_{\rm{es}}$, for Compton scattering
was related to the effective hydrogen number density, $n_{\rm{H}}$, by $n_{\rm{es}}$ = 1.2 $n_{\rm{H}}$.

We used the iron K-shell fluorescence yield of 0.34 \citep{bamby72} and a ratio 17:150 
between the iron $\rm{K}_{\beta}$ and $\rm{K}_{\alpha}$ fluorescence line transition 
probabilities \citep{kikoin76}. Although the iron $\rm{K}_{\alpha}$ fluorescence line consists 
of two components, $\rm{K}_{\alpha1}$ and $\rm{K}_{\alpha2}$ at 6.404 and 6.391 keV, 
respectively (for neutral iron) with a branching ratio of 2:1 \citep{bamby72}, we made no 
distinction between $\rm{K}_{\alpha1}$ and $\rm{K}_{\alpha2}$ photons, adopting a 
common value of 6.40 keV. The iron $\rm{K}_{\beta}$ line is 7.06 keV for neutral iron.

\subsection{Monte Carlo Simulation}

A cross section view of the adopted structure of the torus is illustrated in Figure 2. The center of the torus is 
placed at the origin of the coordinate system, and the equatorial plane of the torus structure lies in the X-Y plane. The structure of 
the torus is described by the following structure parameters: the half-opening angle ${\theta}_{\rm oa}$, the column density $N_{\rm{H}}$ along the equatorial plane, and inner ($r_{\rm{in}}$) and outer ($r_{\rm{out}}$) radii of the 
torus. We assumed $r_{\rm{in}}$/$r_{\rm{out}}$ = 0.01 in our simulation. The angle 
${\theta}_{\rm i}$ in Figure 2 presents the inclination angle of the torus relative to the observer.

A ray-trace method was adopted in our Monte Carlo simulation. The primary X-ray source was 
assumed to emit photons with the energy spectrum, $I(E) \ {\propto} \ E^{-0.9} {\exp}(-E/E_{c})$, where 
$E_{c}$ is a cut-off energy fixed at 360 keV. The spectrum is typical for Type 1 AGNs 
\citep[e.g.,][]{madau94}. 
The primary X-ray source was assumed to be isotropic. 
Each photon from the primary source had both an initial energy and an initial direction of propagation.
In the case that a photon was injected into the torus, an interaction point of the photon was calculated by 
using a random number (see below).  If the Compton scattering occurred at that point, 
the energy and the direction of photon was changed.
The photon was tracked until it escaped the torus structure, or until it was absorbed in the torus.
Note that when the photon with an energy above iron K-edge was absorbed by iron, a K-shell fluorescence 
line was isotropically emitted with a probability of the K-shell fluorescence yield in our simulation, where
the iron absorption event was decided by using the ratio between the photoelectric absorption of iron and 
$\sigma_{\rm abs}$.
Both energies and directions of propagation of all escaping photons were recorded in a photon list.  
For a given observed torus inclination angle of ${\theta}_{\rm i}$ (see Figure 2),  
we extracted photons whose zenith angles of propagation ranged within ${\theta}_{\rm i}$$\pm$
1$^{\circ}$, from the photon list, and then
the extracted photons were accumulated into energy bins to form a spectrum.

The photon transportation in the torus is a key technique in our Monte Carlo simulation. 
The distance $l$ to the next interaction is determined by the probability $p$,
which is described as follows:

\begin{equation}
p = \exp(-{\tau})={\int}_{0}^{l} {\exp}(-{\sigma}_{\rm tot}~ n_{\rm H})dl,
\label{eq:distance_1}
\end{equation}

\noindent
where ${\tau}$ and $\sigma_{\rm{tot}}$ are an optical depth and the total cross section 
of the interaction, respectively. 
The ${\sigma}_{\rm{tot}}$ is comprised of the sum of  ${\sigma}_{\rm{abs}}$ and ${\sigma}_{\rm{es}}$.
By inverting the cumulative probability function, $l$ is expressed as
\begin{equation}
l = \frac{ {\tau} } { n_{\rm{H}} {\sigma}_{\rm{tot} } }= -\frac{1}{n_{\rm{H}}{\sigma}_{\rm{tot}}}{\times}{\rm{ln}}(p).
\label{eq:distance_l}
\end{equation}

\noindent
The distance $l$ is calculated from a uniform random number between 0 
and 1 referred to as $p$.  

Another key in our simulation is Compton scattering.  The scattering angle, ${\theta}_{\rm{scat}}$ (relative to its direction of propagation), is calculated by the differential cross-section for Compton scattering.
We assumed that the differential cross-section was proportional to (1+$\cos^{2}{\theta}_{\rm{scat}}$),
as for the Thomson differential cross-section, since this approximation is efficient for analysis of the $Suzaku$ data
below a few hundreds keV \citep[e.g.,][]{gf91}. We note that  this assumption over-estimates back-scattering relative to forward 
scattering at high energies. The effect of this approximation is seen in the reflection components,
and depends on the geometry and $N_{\rm H}$ of the torus. In order to estimate the effect, we performed a 
simulation in the case of $\theta_{\rm oa}$=40$^{\circ}$, $\theta_{\rm i}$=41$^{\circ}$, and 
$N_{\rm H}$=10$^{25}$ cm$^{-2}$, and found that the change of the reflection components 
due to this assumption was $\sim$10\% at 100~keV. 
%%At 100~keV, the back-scattering Klein-Nishina differential cross-section is $\sim 53\%$ of 
%%the corresponding Thomson approximation. 

The azimuthal angle of the scattering, ${\phi}_{\rm{scat}}$, was randomly selected in the region
 0$^{\circ}$ ${\leq}$ ${\phi}_{\rm{scat}}$ $<$ 360$^{\circ}$. 
The energy of the scattered photon was changed to be $E_{\rm in}$/(1+$\frac{E_{\rm in}}{mc^{2}}$(1$-$cos${\theta}_{\rm{scat}}$ )), where $m$ and $E_{\rm in}$ are electron mass and energy of the photon before Compton scattering, respectively \citep[e.g.,][]{rybicki}. 

%%%%%%%%%%%%%%%%%%%%%%%%%%%%%%%%%%%%%%%%%

%% In this section, we use  the \subsection command to set off
%% a subsection.  \footnote is used to insert a footnote to the text.

%% This section also includes several of the displayed math environments
%% mentioned in the Author Guide.

\section{Results}

\subsection{Simulation Results} \label{bozomath}

Figure 3 is an example of a simulated AGN spectrum with $N_{\rm{H}}$ = $10^{24} \ \rm{cm}^{-2}$, 
${\theta}_{\rm oa}$ = 40$^{\circ}$, and ${\theta}_{\rm i}$ = 45$^{\circ}$. 
We generated a total number of 2.5$\times$10$^{8}$ photons. %%with the spectrum $I(E)$. 
The same number
of photons were generated for each run throughout this paper.
We separated the simulated spectrum into direct and reflection components, where
the direct component has no interaction with the surrounding material, while the reflection component
consists of X-ray photons reflected in the surrounding material.  Since an reflection component, 
which was modeled by $pexrav$ in XSPEC, was required in the $Suzaku$ spectrum of Mrk 3 by
\citet{awaki08}, we divided the simulated reflection component into two, which are referred to as 
the reflection components 1 and 2.  
The reflection component 2 consists of photons emitted from the inner wall of the torus without obscuration by the torus.  The reflection component 1 consists of the rest of 
the reflected photons (see Figure 2).
We have studied the parameter dependence of these three components based on the 
simulations with various $N_{\rm{H}}$, ${\theta}_{\rm oa}$, and ${\theta}_{\rm i}$.

\subsection{Dependence of the Continuum Emission on the Structure Parameters} 

\subsubsection{$N_{\rm{H}}$ Dependence}
For studying the $N_{\rm{H}}$ dependence, we simulated spectra of the three components with 
$N_{\rm{H}}$ = $5{\times}10^{23}, 10^{24}, 2{\times}10^{24}, 3{\times}10^{24}, 5{\times}10^{24}$, 
and $10^{25} \ \rm{cm}^{-2}$, and the simulated components are shown in Figure 4. 
 In these runs, we set ${\theta}_{\rm oa}$ = 40$^{\circ}$ and ${\theta}_{\rm i}$ = 50$^{\circ}$. 
 All the three components, especially the direct component, show a dependence on $N_{\rm{H}}$. 
 It is expected 
that the direct component is affected only by the column density ($N^{\rm ls}$) along our line of sight. 
We compared the simulated direct components with the cut-off power-law models, which 
were affected by both photoelectric absorption and Compton scattering (Figure 5).  
It is found that the models were in good agreement with the simulated direct components. 
Please note that the column density $N^{\rm ls}$ is a function of  
$\theta_{\rm i}$, $\theta_{\rm oa}$ and the ratio $r_{\rm in}/r_{\rm out}$ (=$r$). 
The ratio of $N^{\rm ls}$ to the $N_{\rm H}$ is described as

\begin{equation}
N^{\rm ls}/N_{\rm H}=\frac{r(\cos\theta_{\rm i}-\cos\theta_{\rm oa})+\sin(\theta_{\rm i}-\theta_{\rm oa})}{(1-r)(r\cos\theta_{\rm i}+\sin(\theta_{\rm i}-\theta_{\rm oa})}.
\label{eq:NH}
\end{equation}

\noindent
In the case of $r$=0.01, ${\theta}_{\rm oa}$ = 40$^{\circ}$, and
${\theta}_{\rm i}$ = 50$^{\circ}$, for example, the value of $N^{\rm ls}/N_{\rm H}$ is deduced to be 0.97.

\subsubsection{Dependence on the Half-opening Angle, ${\theta}_{\rm oa}$}

Three components depended on the half-opening angle, ${\theta}_{\rm oa}$, but we discuss the  
 ${\theta}_{\rm oa}$-dependence of only the reflection components 1 and 2, since the 
 ${\theta}_{\rm oa}$-dependence of the direct component, which is affected by $N^{ls}$, has been
 described by equation 3.

The left panel in Figure 6 shows simulated spectra of the reflection component 1 for ${\theta}_{\rm oa}$ = 
$10^{\circ} - 70^{\circ}$ in steps of 10$^{\circ}$.  The values of $N_{\rm{H}}$ and ${\theta}_{\rm i}$ were
fixed at $10^{24} \ \rm{cm}^{-2}$ and $90^{\circ}$ in these runs, respectively. 
The spectral shape of the reflection component 1 has a small dependence on ${\theta}_{\rm oa}$, and the 
total photon count 
in the reflection component 1 decreases with increasing $\theta_{\rm oa}$.   We plot the total count 
 in Figure 7.  Since 
some photons emitted from the central source escape through the opening area of the torus, the intensity of 
the reflection component 1 is dependent upon ${\theta}_{\rm oa}$.  We compared the total counts of this 
component with the covering fraction of the torus,  $\cos \theta_{\rm oa}$.  In the right panel in Figure 8, 
we plot the normalized total counts, which are divided by the total counts in the case of spherical
distribution of the surrounding material.   It is found 
that the ${\theta}_{\rm oa}$-dependence of the reflection component 1 can be roughly explained 
by the covering fraction of the surrounding material. 
For large $N_{\rm H}$ of 10$^{25}$ cm$^{-2}$, the total counts do not
follow this $\theta_{\rm oa}$-dependence as for low $N_{\rm H}$ due to the absorption of the reflection component
by the torus itself.

The right panel in Figure 6 shows the ${\theta}_{\rm oa}$-dependence of the reflection component 2. 
In these runs, we set ${\theta}_{\rm i}$=${\theta}_{\rm oa}$+1$^{\circ}$ in order to obtain a high intensity 
of the reflection component 2 under the condition that the nucleus is obscured by the torus.
The total count of the 
reflection component 2 are shown in Figure 7. It is found that
the reflection component 2 is to be zero at both ${\theta}_{\rm oa}$ =0 and 90$^{\circ}$, and has a
maximum intensity at ${\theta}_{\rm oa} \ {\sim} \ 30^{\circ}$. 
The reflection component 2 is emitted from the visible inner wall of the torus. 
Thus, this component vanishes at ${\theta}_{\rm oa}$ =0 due to there being no inner wall.
This component is also associated with the total number of photons injected into the torus from the central
source, which is related to the solid angle of the torus.   Therefore,
the intensity of the reflection component 2 is associated with a combination of both apparent 
size of the visible inner wall of the torus and solid angle of the torus, i.e.
$\cos \theta_{\rm oa} \cos (2\theta_{\rm oa}-\pi/2)$.  The left panel in Figure 8 shows the total counts of the 
reflection component 2 in $N_{\rm H}$=10$^{23}$, 10$^{24}$, and 10$^{25}$ cm$^{-2}$.
For the case of $N_{\rm H}$=10$^{25}$ cm$^{-2}$, the $\theta_{\rm oa}$-dependence is well 
reproduced by the combination described as  $\cos \theta_{\rm oa} \cos (2\theta_{\rm oa}-\pi/2)$, 
since the reflection occurs mainly on the inner surface of the torus.

\subsubsection{Dependence on the Inclination Angle ${\theta}_{\rm i}$}

We show simulated spectra of the reflection components for ${\theta}_{\rm oa}$=10$^{\circ}$ with 
${\theta}_{\rm i}$=$11^{\circ}$ to 
$71^{\circ}$ in steps of 20$^{\circ}$ in  Figure 9, and show the total counts at 
${\theta}_{\rm oa}$=10$^{\circ}$ and 30$^{\circ}$ against ${\theta}_{\rm i}$ from 1$^{\circ}$ to 89$^{\circ}$ 
in steps of 2$^{\circ}$ in Figure 10. In these runs,  $N_{\rm{H}}$ was fixed at $10^{24} \ \rm{cm}^{-2}$. 
The reflection component 1 has a weak ${\theta}_{\rm i}$-dependence (see Figure 9), and has a 
similar intensity in ${\theta}_{\rm i}$ $>$ ${\theta}_{\rm oa}$ +10$^{\circ}$ (Figure 10). The 
${\theta}_{\rm i}$-dependence is consistent with the fact that the reflection component 1 depends on the 
covering factor of the surrounding material as mentioned in the previous subsection.
The unified model of Seyfert galaxies predicts that ${\theta}_{\rm i}$ is larger than ${\theta}_{\rm oa}$
in Seyfert 2 galaxies.  Thus, for most Seyfert 2 galaxies, the reflection component 1 should show a 
small ${\theta}_{\rm i}$-dependence, and this component should depend on mainly $\theta_{\rm oa}$ and  
$N_{\rm H}$ of the torus. In the case of $\theta_{\rm oa}$ $<$ $\theta_{\rm i}$ $<$ $\theta_{\rm oa}$+10$^{\circ}$,
the reflection component 1 should have a strong $\theta_{\rm i}$-dependence. 

The reflection component 2 shows a strong ${\theta}_{\rm i}$-dependence (Figure 9), because the component 
is emitted from the visible inner wall, whose area is decreasing with increasing ${\theta}_{\rm i}$.  
If we observe the source edge-on, i.e.; $\theta_{\rm i}$
=90$^{\circ}$, the area of the wall is apparently zero for the observer. On the other hand, if we observe the 
source face-on, i.e.; $\theta_{\rm i}$=0$^{\circ}$, the projected area (i.e. the area projected onto the sky,
perpendicular to the line of sight) is largest.  The $\theta_{\rm i}$-dependence of the reflection component 
2 is roughly explained only by that of the projected area. 
Note that larger portion of the projected area, or the visible inner region, is obscured by 
the near side of the torus with increasing $\theta_{\rm i}$ in $\theta_{\rm i}$ $>$ $\theta_{\rm oa}$.
The spectral shape of the reflection component 2 is also affected by ${\theta}_{\rm i}$ as shown in Figure 9.
With increasing $\theta_{\rm i}$, the low-energy cut off becomes more pronounced. This is because 
only X-rays reflected from the visible inner region of the torus (see Figure 2) contribute to the reflection 
component 2, as defined. 
Some photons emitted from the central source enter the far side
of the torus and Compton scattered. A part of the scattered photons 
are absorbed inside the torus before they ever reach the visible 
inner region. Since the length of the path of such photons increases 
with $\theta_{\rm i}$, more photons are absorbed before they reach
the visible region.
%%Some X-rays injected from the central source into the torus are absorbed before 
%%they reach the visible inner region. For increasing $\theta_{\rm i}$, the distance to the visible region 
%% from the central source is increasing, and  more of  the soft X-rays are absorbed before reflection. 
 As a result, we can see a ${\theta}_{\rm i}$-dependence on the spectral shape.

\subsubsection{Note on dependence on $r_{\rm{in}}$/$r_{\rm{out}}$}

The ratio $r_{\rm in}/r_{\rm out}$ (=$r$) was fixed at 0.01 in our simulation. We briefly discuss 
the spectral dependence on this ratio.  
We simulated the reflection components with different values of the ratio, 0.001, 0.01, 0.1, and 1.  In these 
runs, $N_{\rm{H}}$,   ${\theta}_{\rm oa}$, and ${\theta}_{\rm i}$ were set to be $10^{24} \ \rm{cm}^{-2}$, 
 40$^{\circ}$, and $41^{\circ}$, respectively.
The reflection component 1 shows a very weak dependence on this ratio,
since these simulations were carried out for the same ${\theta}_{\rm oa}$, and this component depends
mainly on the covering fraction of the torus.  
On the other hand, the reflection component 2 below $\sim$ 4 keV depends on the ratio $r$
(Figure 11).  Reflections occur more frequently than absorption at the inner 
region of the torus.  Thus, in the case of the small $r$,
 the closest region to the source is obscured by the torus, since we consider the case of Seyfert 2 galaxies,
  ${\theta}_{\rm i}$ $>$ ${\theta}_{\rm oa}$.  This effect is also seen in the ${\theta}_{\rm i}$ dependence
  of the reflection component 2 (see Figure 9).  
  This study indicates that it is difficult to produce a strong unabsorbed reflection-component in our 
simple torus geometry with a small $r$ (=0.001).

\subsection{X-ray luminosity absorbed by the torus}

In our simulation, X-rays emitted from the central source were absorbed by the dusty torus. 
%%It is useful to investigate the X-ray luminosity absorbed by the torus for understanding the correlation 
%%between infrared and X-ray luminosites of AGNs, since the dusty torus is heated by optical, UV, and 
%%X-ray photons from the central source. 
It is important to compare the absorbed X-ray luminosity with the infrared luminosity, since  
the infrared luminosity is a good indicator for the absorbing luminosity in the torus.
We estimated the fraction of the absorbed luminosity with respect to the intrinsic 
source luminosity.
The intrinsic luminosity of the central X-ray source was calculated from the assumed spectrum in 
section 2.2, and the absorbed X-ray luminosity was derived from the subtraction of an output luminosity from 
the intrinsic luminosity, where the output luminosity was deduced from a total energy of the escaping 
photons from the torus structure.
 Figure 12 shows the absorbing fraction to the 1--100 keV intrinsic luminosity as a function of 
 $\theta_{\rm oa}$ in $N_{\rm H}$=10$^{22}$, 10$^{23}$, 10$^{24}$, and 10$^{25}$ cm$^{-2}$. The 
 dependence of the absorbing fraction is roughly explained by $\cos$ $\theta_{\rm oa}$, which displays the 
 covering fraction of the dusty torus. In the case of $N_{\rm H}$ = 1$\times$10$^{24}$ cm$^{-2}$ and 
 $\theta_{\rm oa}$=45$^{\circ}$, the absorbing fraction of 0.37 is obtained, and the absorbed luminosity is estimated to 
 be 1.1$\times$10$^{43}$ erg s$^{-1}$ for the 1--100 keV intrinsic luminosity of 3$\times$10$^{43}$ erg 
 s$^{-1}$, which corresponds to the 2--10 keV intrinsic luminosity of 1$\times$10$^{43}$ erg s$^{-1}$.

The infrared luminosity has a good correlation with the 2--10 keV luminosity. By using the relation between
these luminosities by Mulchaey et al. (1994), the infrared luminosity for an AGN with the 2--10 keV X-ray luminosity of 1$\times$10$^{43}$ 
erg s$^{-1}$ is
 estimated to be 10$^{44}$ erg s$^{-1}$, which is about 10 times larger than the absorbed X-ray luminosity.
The dust in the torus is heated by optical and UV photons as well as X-ray photons. Our study shows an estimation of the fraction of the absorbed X-ray luminosity to dust heating.
We indicate that the absorbed luminosity depends on the geometry of the torus, $\theta_{\rm oa}$ in 
Figure 12. Lutz et al. (2004) pointed out that 
the scatter in the relation between mid-infrared and absorption corrected hard X-ray luminosities was about 
one order of magnitude, and that the scatter was likely caused by the geometry of the absorbing dust. 
The scatter seen in the relation may be explained by the $\theta_{\rm oa}$-dependence of the absorbed 
luminosity.

\subsection{Dependence of the Iron-line Emission on the Structure Parameters} 
A prominent iron line with an equivalent width of $>$ a few 100 eV is an important characteristic 
of Seyfert 2 galaxies. The dependence of the equivalent width of the iron line on the structure parameters of the 
torus have been studied in previous work \citep[e.g.,][]{awaki91, ghi94, levenson02}.  
These studies were mainly
performed for the equivalent width relative to the total continuum emission, comprised of the
sum of the direct and reflection components.  In our study, we have 
investigated the equivalent width relative to the reflection component (hereafter $EW_{\rm ref}$) as 
well as the equivalent width to the total continuum emission (hereafter $EW_{\rm tot}$).  In Figures
13 and 14, we plot the equivalent width as a function of $N_{\rm H}$ and $\theta_{\rm oa}$
respectively (note that for $\theta_{\rm oa}$=0 we use a spherical distribution instead of
using our simple torus model with $\theta_{\rm oa}$=0).

The left panel in Figure 13 shows the $N_{\rm H}$-dependence of $EW_{\rm tot}$ fixed at 
$\theta_{\rm oa}$=30$^{\circ}$. We found that the $N_{\rm H}$-dependence is similar to
those of previous studies, although our results are about 1.7 times 
larger than those obtained by \citet{ghi94}.  The difference may be caused by the difference
of iron abundance used in the simulations.  The right panel in Figure 13 shows the $\theta_{\rm oa}$-dependence of $EW_{\rm tot}$. 
%In Figure 13, we plot the equivalent width for the spherical distribution at $\theta_{oa}$=0, instead of
%$\theta_{oa}$=0 in our simple torus model.
The $EW_{\rm tot}$ in the Compton-thin region ( $N_{\rm H}$ $<$ 10$^{24}$ cm$^{-2}$) 
decreases with increasing $\theta_{\rm oa}$. This 
$\theta_{\rm oa}$-dependence is represented by a function of $\cos \theta_{\rm oa}$, similar to the
reflection component 1, since the iron line intensity is proportional to the
solid angle subtended by the surrounding matter at the source. 

The left panel in Figure 14 shows the $N_{\rm H}$-dependence of $EW_{\rm ref}$ for
$\theta_{oa}$=30$^{\circ}$. The $EW_{\rm ref}$ shows little $N_{\rm H}$ dependence in the 
$N_{\rm H}$ $<$10$^{24}$ cm$^{-2}$, since the $EW_{\rm ref}$ is mainly 
determined by the ratio 
between the absorption and scattering cross-sections. In the Compton-thick region, 
the $EW_{\rm ref}$ shows large $\theta_{\rm oa}$- and $\theta_{\rm i}$-dependences.
This is caused by the fact that the contribution of the reflection component 1 to the reflection continuum 
at the iron band is decreasing with increasing $N_{\rm H}$ (see Figure 4).

We noted that the $EW_{\rm ref}$ in the Compton-thick regime is
greater than 1000 eV in our simulations. Thus, observed lower values
of $EW_{\rm ref}$, less than 1000 eV, may indicate a low
metal abundance of iron.

%%%%%%%%%%%%%%%%%%%%%%%%%%%%%%%%%%%%%%%%%

\section{Application to Observed Spectrum}

By means of Monte Carlo simulations, we showed that the three continuum components depend on the 
structure parameters of the torus, $N_{\rm{H}}$, $\theta_{\rm oa}$, and $\theta_{\rm i}$. Our study suggests  
that we can estimate the structure of the torus by determining the three components in an observed 
spectrum. We therefore made a new model for spectral fitting based on our simulations with 1$\times$
10$^{9}$ photons in each run.  
In our new model, the direct component is reproduced by $phabs*compcabs*(cutoffpl)$ in 
$XSPEC$, where $compcabs$ is a new model that we made for representing Compton scattering, 
and the two reflection components are reproduced by using table models, which
have parameters of photon index,   $N_{\rm{H}}$, ${\theta}_{\rm oa}$, and $\theta_{\rm i}$, since the
reflection components are too complex to be represented by numerical expression. The table 
models cover the ranges of photon index of $1.5-2.5$, $N_{\rm{H}}$ of
 $10^{22}-10^{25} \ \rm{cm}^{-2}$, ${\theta}_{\rm oa}$ of $0^{\circ}-70^{\circ}$, and $\theta_{\rm i}$ of 
 $0-90^{\circ}$. The details of the parameters of the table models are listed in Table 1.

The Suzaku satellite can obtain a wide-band spectrum with good quality \citep{mitsuda07}.  We
 applied the new model to the wide-band spectrum of Mrk 3 observed by the Suzaku satellite on 2005 October 22-23 during the SWG phase. 
We obtained XIS and HXD spectra in the same manner as described by \citet{awaki08}, and 
then simultaneously fitted the XIS and HXD spectra above 1 keV with the new model defined as follows:
\begin{equation}
I({\rm{ph \ s^{-1} \ cm^{-2} \ keV^{-1}}}) = PL1+ e^{-{\sigma}_{\rm a}N_{\rm{H1}}}e^{-{\sigma}_{\rm{es}}N_{\rm{H1}}}PL2+reflection1(N_{\rm EL})+reflection2(N_{\rm H2})+ELs,
\label{eq:baseline_model}
\end{equation}
where PL1 and PL2 are power law components with a high energy cut-off.
The high energy cutoffs of both PL1 and PL2 were fixed at 360 keV, which is consistent with 
the cut-off energy ($E_{cut}$) of the power law component $E_{cut}$ $>$ 200 keV obtained by 
\citet{cappi99}.  $N_{\rm{H1}}$ is the column density along our line of sight. We used the absorption cross-section of $vern$ for  ${\sigma}_{\rm a}$
 in the $phabs$ model in $XSPEC$ v12.4. The abundances of \citet{ag89} were used. The ${\sigma}_{\rm{es}}$ is a Compton scattering cross-section, which is used in the $compcabs$ model.  
 %%We made a new model $compcabs$ in $XSPEC$ for representing Compton scattering. 
%%The $compcabs$ is the same as the $cabs$ in $XSPEC$ except for the cross-section of the scattering.
The reflection components 1 and 2 in our simulation were reproduced by the two table models, 
$reflection1$ and $reflection2$, respectively.  We set their photon indices equal to that of PL2,
while we did not link their nomalizations to that of PL2, and their column densities along the equatorial plane
($N_{\rm H2}$) were not linked to the column density ($N_{\rm H1}$) of PL2 in our spectral fit. 
The emission lines (ELs) seen in the spectrum were represent by the sum of Gaussian components:
\begin{equation}
ELs=\sum_{\rm i} gauss(E_{\rm i}, {\sigma}_{\rm i}, N_{\rm i}),
\label{eq:baseline_model1}
\end{equation}
where $E_{\rm i}$, ${\sigma}_{\rm i}$, and $N_{\rm i}$ are the center energy, width, and intensity of the $i$-th line. We fixed $E_{\rm i}$ and ${\sigma}_{i}$ at the values obtained by \citet{awaki08}.

We fitted the spectra with the new table models in the energy range from 1 to 70 keV. 
Since the table models were generated from simulations, the table models have statistical deviations.
A typical standard deviation per 20 eV bin of  the sum of $reflection$ 1 and 2 in
the 3--5 keV band is about 2\%, which is about 1/5 of the statistical error of the observed data in this
energy band. Due to the deviations of the table models, the $\chi^{2}$ will have a fluctuation of 
about (bin number)$\times (\frac{1}{5})^{2}$, which is estimated to be about 5.
Thus, it is hard to find the best-fit parameters and their confidence regions with the $\chi^{2}$-fitting 
procedure. We performed $\chi^{2}$-test with our table models on the 
parameter grids of $\theta_{\rm oa}$ and $\theta_{\rm i}$ in order to examine whether we can obtain the 
structure parameters from the spectral fit with the table models. 
Table 2 lists the $\chi^{2}$ on the grids with 612 d.o.f.  
The minimum $\chi^{2}$ ($\chi^{2}_{\rm min}$) was obtained for $\theta_{\rm i}$ $<$ $\theta_{\rm oa}$, 
and the $\chi^{2}$ was more than $\chi^{2}_{\rm min}$ + 30 for 
$\theta_{\rm i}$ $>$ $\theta_{\rm oa}$+3$^{\circ}$. Since the $\theta_{\rm i}$ must be larger than the 
$\theta_{\rm oa}$ in our simple torus model (otherwise the $N_{\rm H1}$ becomes zero), the $\chi^{2}$ 
study indicates that we observed the Mrk 3 nucleus near the edge of the torus.
This is expected from the strong unabsorbed reflection component of Mrk 3 in the baseline model.

On the other hand, we found that it is difficult to constrain the half-opening angle  $\theta_{\rm oa}$ 
from the spectral analysis, since the shapes of the reflection components 1 and 2 show a little 
$\theta_{\rm oa}$-dependence (Figure 6).  Although their intensities depend on $\theta_{\rm oa}$ as shown in
Figure 7,
the intensities of the reflection components also depend on the luminosity of the central source.
Figure 15 shows the change of the normalization of the direct and reflection
components for $\theta_{\rm i}$=$\theta_{\rm oa}$+1$^{\circ}$.  The normalization of the reflection 
component is roughly proportional 1/$\cos \theta_{\rm oa}$. This relation indicates that the contribution 
of the reflection components to the observed spectrum is nearly constant.   
Furthermore, the spectral shape of the reflection component 1 in the 5--70 keV band is similar to 
that of the direct component 
at $N_{\rm H} \sim$10$^{24}$ cm$^{-2}$ (see Figure 3).  As a result, it is hard to constrain
$\theta_{\rm oa}$, even if we link the intensities between the direct and
reflection components.  We note that the iron EW is not helpful to find the $\theta_{\rm oa}$ of Mrk 3,
since $EW_{\rm ref}$ has a little dependence on $\theta_{\rm oa}$ in $N_{\rm H}$ $<$ 10$^{24}$
cm$^{-2}$.
 
%% NH issue 
In order to constrain $N_{\rm H2}$, we fitted the observed spectrum with our model on the grid
of $N_{\rm H2}$.  Since we did not constrain the $\theta_{\rm oa}$ from the spectral fit, we
fixed  $\theta_{\rm oa}$ at 50$^{\circ}$, and $\theta_{i}$ at $\theta_{\rm oa}$+1$^{\circ}$.
The value of $N_{\rm H1}$ was a free parameter in this fit. 
We found that there is a $\chi^{2}$ minimum around 
$\sim$10$^{24}$ cm$^{-2}$, and that the $\chi^{2}$ is greater than $\chi^{2}_{\rm min}$+30 for
$N_{\rm H2}$ $<$ 6$\times$10$^{23}$ and for $N_{\rm H2}$ $>$ 2$\times$10$^{24}$ cm$^{-2}$.
 We found that we constrain the structure parameter $N_{\rm H}$ of the torus by using
 our new model. 
 
%% real model
The opening angle may be estimated from the opening angle of the NLR. \citet{capetti95} found a NLR opening angle in Mrk 3 of $>$ $100^{\circ}$.  The half opening angle of $>$$50^{\circ}$ is larger than 
the estimation by \citet{ruiz01} due to the inclusion of all the Z-shape emission components in the NLR in the estimation of  \citet{capetti95}.
We here set $\theta_{oa}$ = $50^{\circ}$ and $\theta_{\rm i}$ = $51^{\circ}$ in our spectral fit. Furthermore,
the column density $N_{\rm H1}$ was linked with $N_{\rm H2}$ by
 $N_{\rm H1}$=0.74 $N_{\rm H2}$, by using the equation (3).
We obtained a small reduced $\chi^{2}$ value of 1.18 (d.o.f.=613), which is comparable to that with the
baseline model. Table 3 shows the best-fit parameters, and Figure 16 shows the best-fit spectrum.
%% gamma
The photon index of the power law component and the column density of $N_{\rm H1}$ were 
estimated to be $\sim$1.82 and $\sim$1.1$\times$10$^{24}$ cm$^{-2}$, respectively.
%% flux and luminosity 
The intrinsic luminosity of the power law component in the 2--10 keV band was estimated to be 
2.1$\times$10$^{43}$ erg s$^{-1}$, which is about 1.3 times that of the estimate with the baseline 
model and is consistent with that obtained by using the model with $plcabs$ \citep{awaki08}, 
which describes the X-ray transmission, correctly taking into account Compton scattering 
\citep{yaqoob97}.  On the other hand, the intrinsic luminosity 
irradiating the accretion torus is estimated to be 5.1$\times$10$^{43}$ erg s$^{-1}$ from the
normalization of the reflection component.  
The discrepancy of the intrinsic luminosities between the direct and reflection 
components may be arisen by a time lag of the reflection component, since the long time 
variability of Mrk 3 has been reported and the distance of the accretion torus is estimated to be greater than
1 pc from the center \citep[e.g.,][]{awaki00, awaki08}.

%%Although the normalization of the reflection component is consistent with that of the direct component
%%within 90\% error region, 
%%The ratio of the 2--10 keV to [O$_{\rm III}$] $\lambda$5007 luminosities for Mrk 3 is about 2.3, which
%%is consistent with those for a sample of Seyfert 1 galaxies (Heckman et al. 2005).

\section{Summary and Conclusion}

We simulated AGN spectra by using the ray-trace method, and made a new model for  
fitting spectra of Compton-thick AGNs.
In our simulations we assumed an accretion torus surrounding a nucleus, which was 
characterized by $\theta_{\rm oa}$, $\theta_{\rm i}$, $N_{\rm H}$, and the ratio of the inner and outer 
radii of the torus.
We considered interactions of photoelectric absorption, iron fluorescence, and Compton scattering in 
the simulation.  The simulated spectra
were separated into three components: one direct component and two reflection components. 

The direct component consists of X-ray photons which have no interaction with the surrounding 
material, and the component
is only affected by column density along our line of sight. In fact, the direct component was well modeled 
with a cut-off power law emission affected by both photoelectric absorption and Compton scattering.
On the other hand, the reflection components had not only an $N_{\rm H}$-dependence but also both
$\theta_{\rm oa}$- and $\theta_{\rm i}$-dependences. The reflection component 1 shows a $\theta_{\rm oa}$-dependence, which is explained by the covering factor of the torus.
The reflection component 2 has dependence on both $\theta_{\rm oa}$ and $\theta_{\rm i}$. 
The dependence of the reflection component 2 is roughly explained by the apparent size of the visible inner wall of the torus.  

We fitted the $Suzaku$ Mrk 3 spectrum with the new model based on our simulations, and found that the spectrum 
could be represented by our model.  The structure parameters of the torus of Mrk 3 were estimated 
with the new model: $\theta_{\rm i}\sim \theta_{\rm oa}+1^{\circ}$ and $N_{\rm{H}} \sim 10^{24} \ \rm{cm}^{-2}$,
although it was hard to constrain $\theta_{\rm oa}$ from our spectral analysis.
We estimated the intrinsic luminosity of the direct component and the intrinsic luminosity irradiating the
surrounding matter.  Assuming $\theta_{\rm oa}$=50$^{\circ}$ and $N_{\rm H1}$=0.76$N_{\rm H2}$,
the 2--10 keV luminosity of the direct component was estimated to be about 2/5 of that irradiating 
the surrounding matter.  This may be explained by time variability of Mrk 3 and time lag between
the direct and reflected lights.

We demonstrated that we can bring out the structure of the torus from an observed X-ray spectrum 
with our new model.  The wide-band X-ray spectra will be helpful to determine the structure of AGNs.

\acknowledgments

We also thank Drs. T. Yaqoob and K. Murphy for useful discussions and careful reading. We also thank 
the anonymous referee for helpful comments and suggestions. 
This study is carried out in part by the Grant support for Scientific Research of Ehime university (H.A.) 
and the Grant-in-Aid for Scientific Research  (17740124 Y.T.) of the Ministry of Education, Culture, 
  Sports, Science and Technology.

%Figure ------

\clearpage

\begin{figure}[!h]
\includegraphics[angle=0,scale=.32]{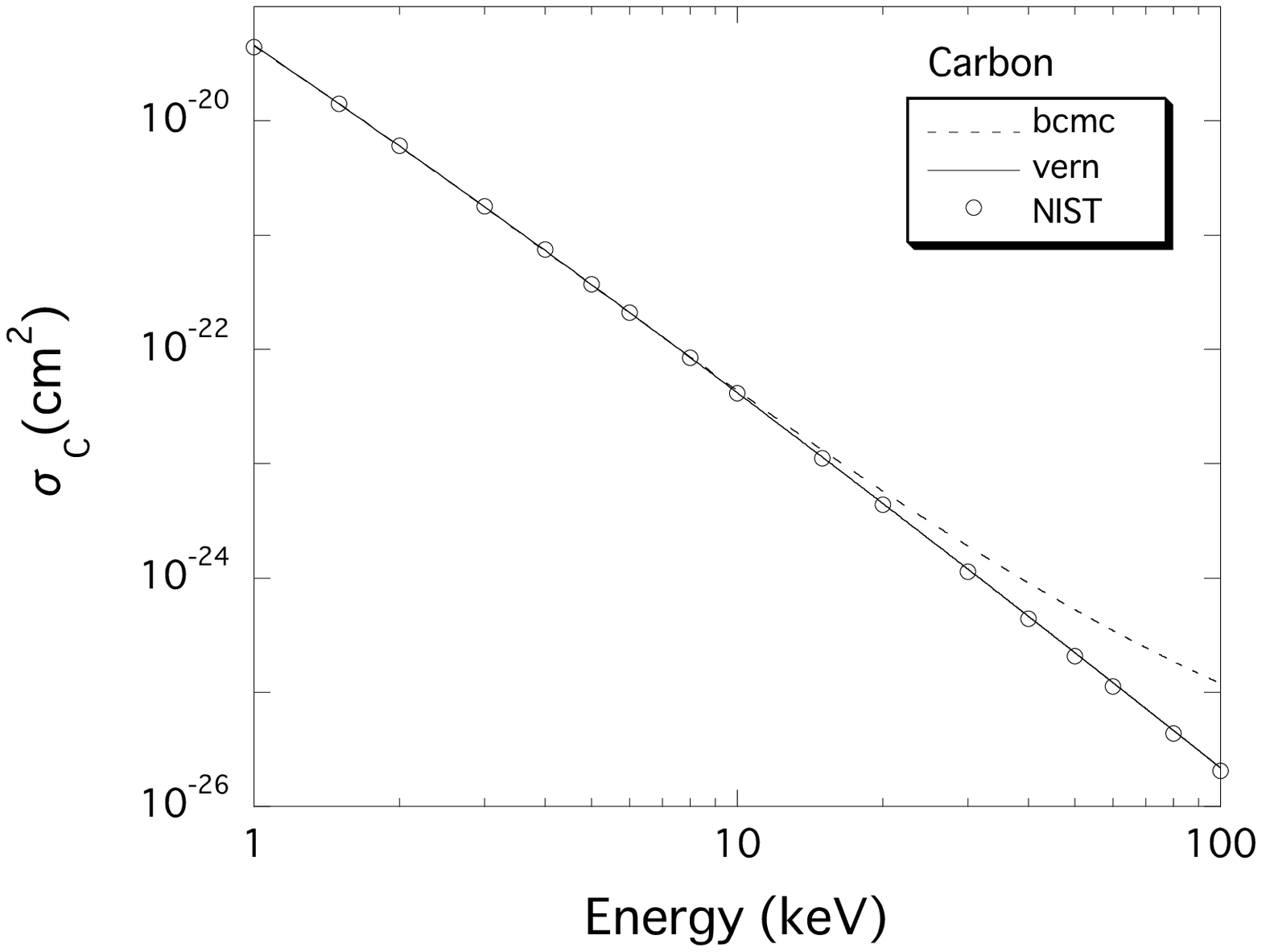}
\includegraphics[angle=0,scale=.32]{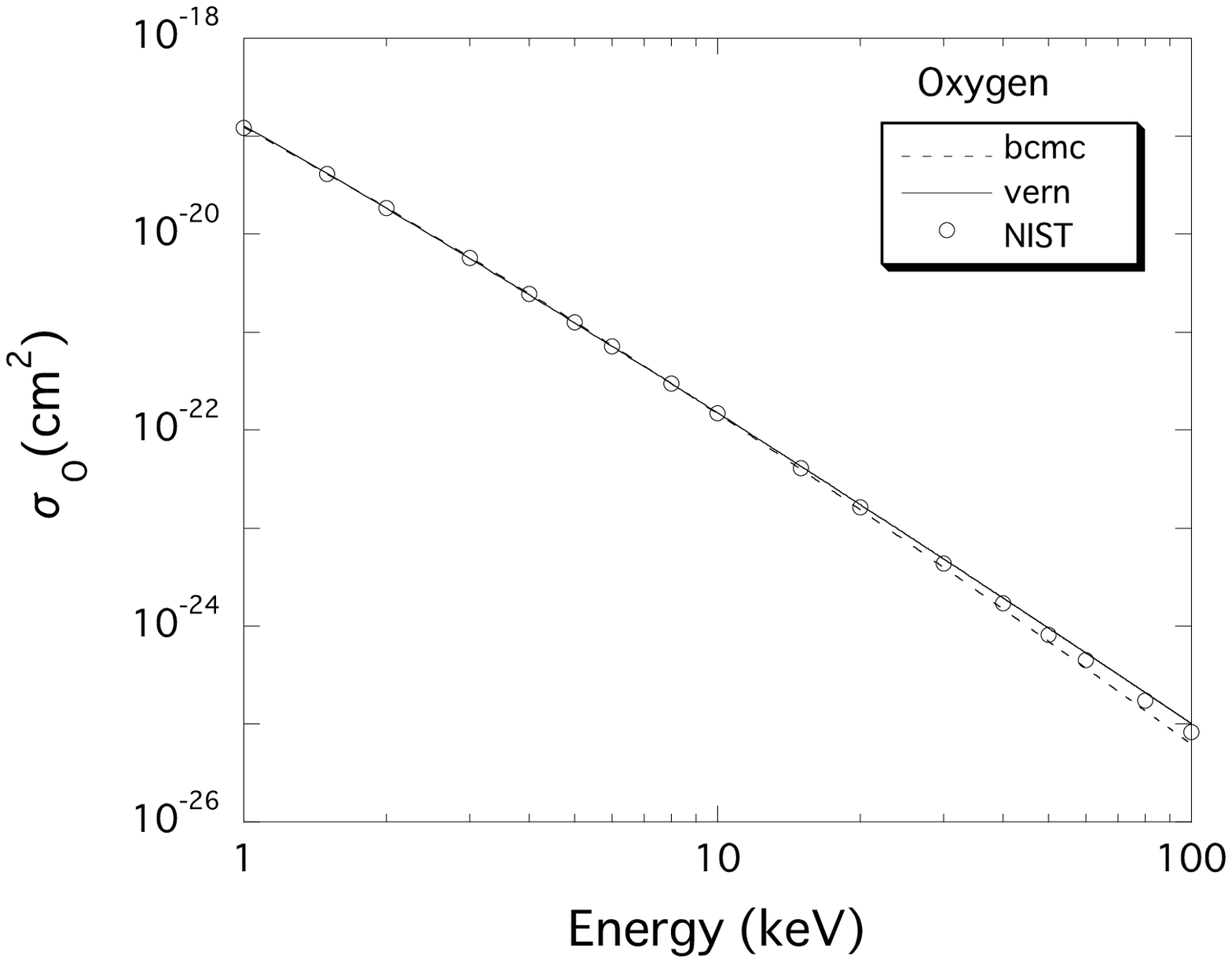}
\includegraphics[angle=0,scale=.32]{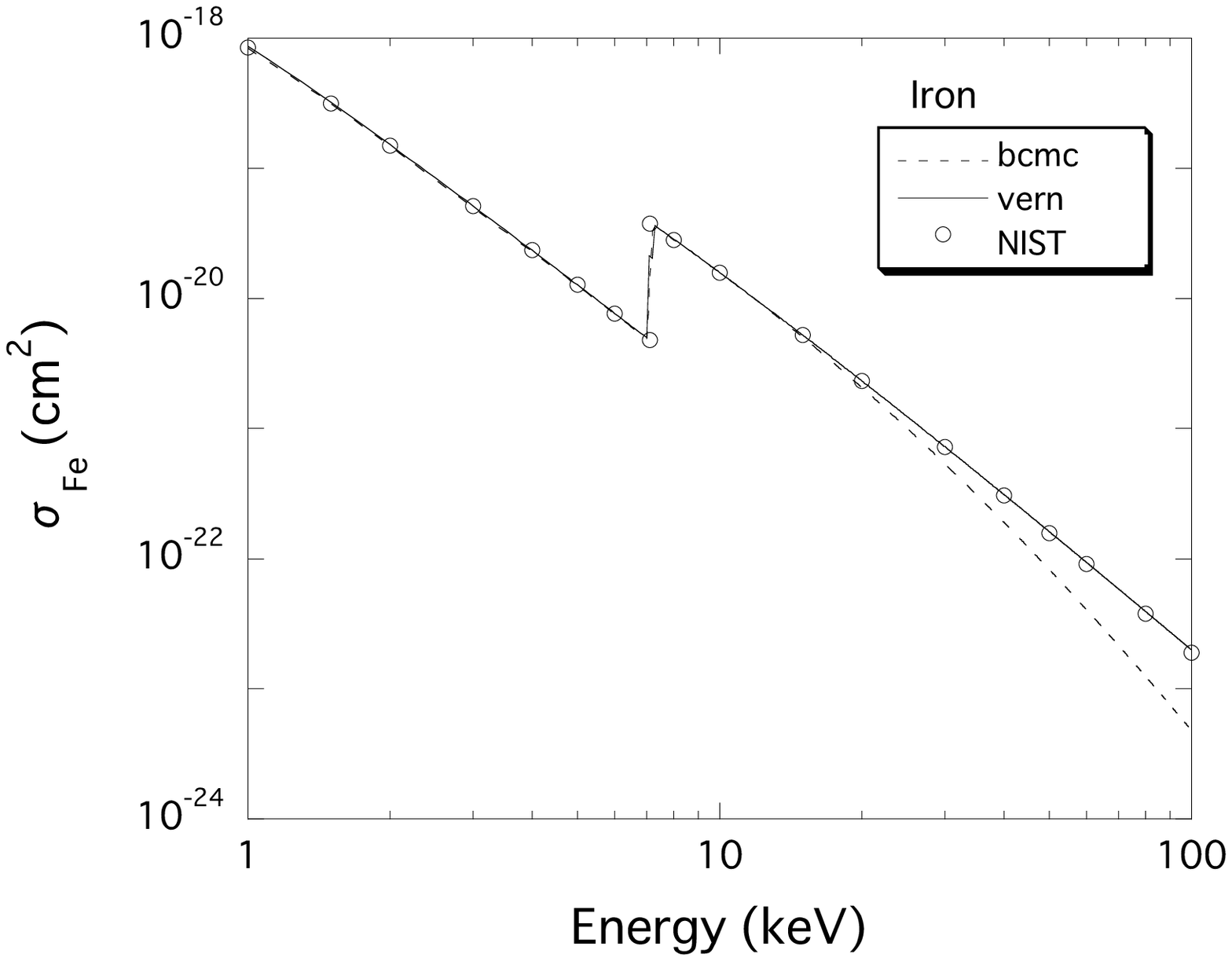}
\caption{Photoabsorption cross-section of Carbon, Oxygen, and Iron. The open circles
indicate those obtained from the NIST database. The solid and dashed lines show those
by \citet{vern96} and \citet{bm92}, respectively.
Their cross-sections are referred to as $vern$ and $bcmc$ in $XSPEC$.  
}
\end{figure}

% structure image
\clearpage

\begin{figure}[!h]
\includegraphics[angle=0,scale=1.0]{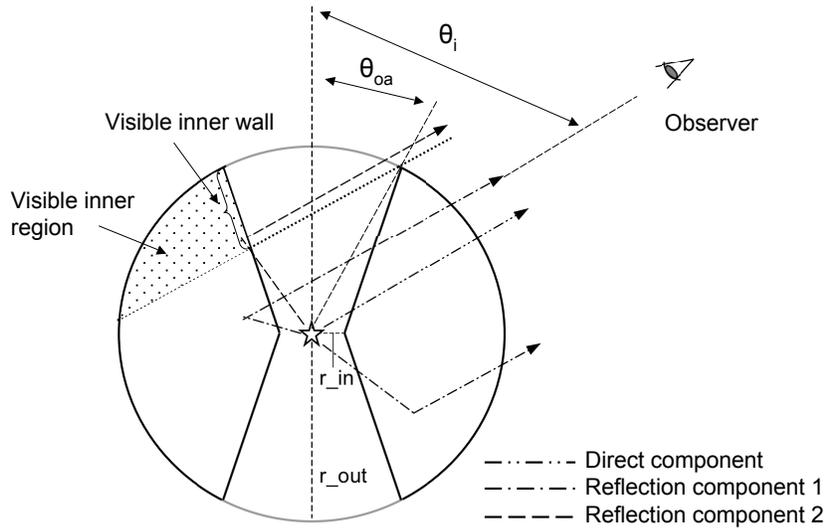}
\caption{A cross section view of the torus structure in our simulation. The primary X-ray source is located at the center of an obscuring torus.  The torus structure is characterized by the half-opening angle ${\theta}_{\rm oa}$,
the inclination angle of the torus from an observer ${\theta}_{\rm i}$, the column density $N_{\rm{H}}$ 
along the equatorial plane, and the ratio of $r_{\rm{in}}$ to $r_{\rm{out}}$.   
A simulated spectrum is separated into three components: one direct component and two reflection components. The two reflection components are referred as reflection component 1 and 2. 
The reflection component 2 consists of reflection light from the visible inner wall of the torus.
The reflection component 1 consists of the rest of the reflection light.
}
\end{figure}

% sample result

\begin{figure}[!h]
\includegraphics[angle=270,scale=.38]{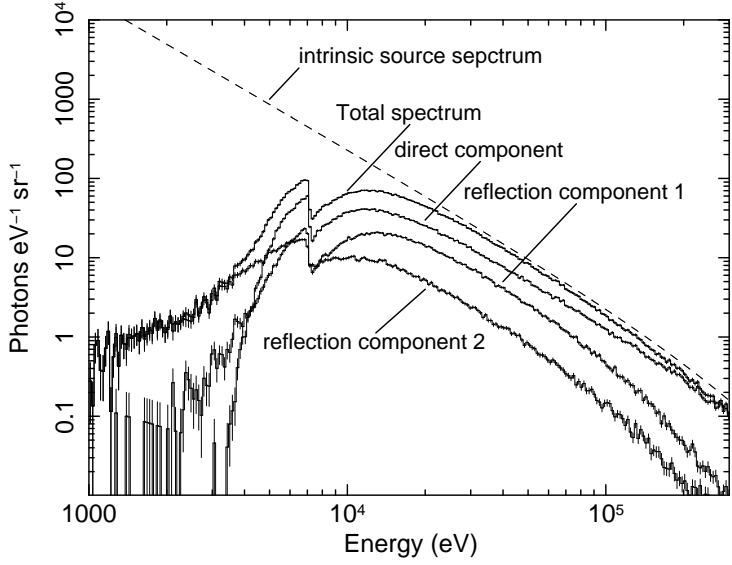}
\caption{An example of our simulated AGN spectrum with $N_{\rm{H}}$ = $10^{24} \ \rm{cm}^{-2}$, 
 ${\theta}_{\rm oa}$ = 40$^{\circ}$, and ${\theta}_{\rm i}$ = 45$^{\circ}$.  
 The simulated spectrum is divided into three components : the direct component, the 
 reflection component 1, and the reflection component 2 as shown in this figure. The dashed line
 displays the intrinsic source spectrum.}
\end{figure}

%column density

\begin{figure}[!h]
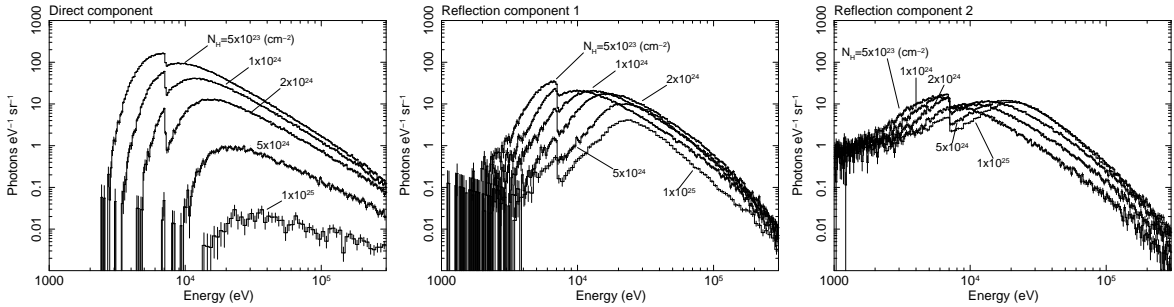

\includegraphics[angle=270,scale=.2]{f4a.eps}
\includegraphics[angle=270,scale=.2]{f4b.eps}
\includegraphics[angle=270,scale=.2]{f4c.eps}
\caption{$N_{\rm{H}}$ dependence of the three components for ${\theta}_{\rm oa}$ = 40$^{\circ}$ and 
${\theta}_{\rm i}$ = 45$^{\circ}$. The column density has values of $N_{\rm{H}}$ = $5{\times}10^{23}, \ 10^{24}, \ 2{\times}10^{24}, \ 5{\times}10^{24} \ \rm{and} \ 10^{25} \ \rm{cm}^{-2}$. }
\end{figure}

%compcabs

\begin{figure}[!h]
\includegraphics[angle=0,scale=.5]{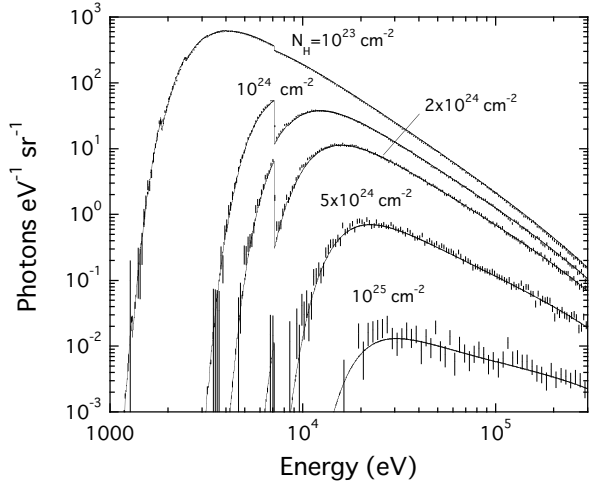}
\caption{Comparison between the simulated direct component and the cut-off power law model affected by 
both photoelectric absorption and Compton scattering. }
\end{figure}

%half-openingangle

\begin{figure}[!h]
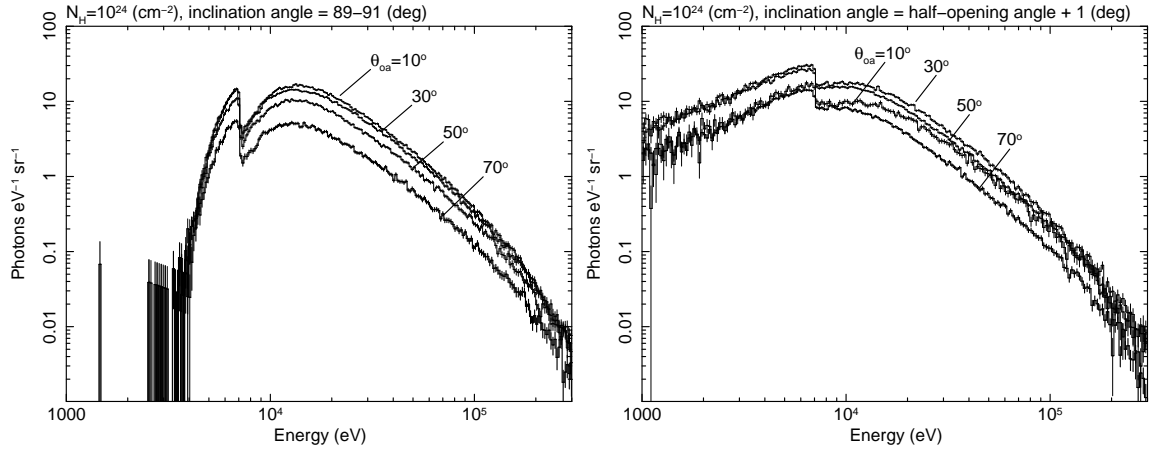

\includegraphics[angle=270,scale=.30]{f6a.eps}
\includegraphics[angle=270,scale=.30]{f6b.eps}
\caption{ ${\theta}_{oa}$-dependence of the reflection components 1 and 2. For the simulations
of the reflection component 1,
$N_{\rm{H}}$ and ${\theta}_{\rm i}$ were fixed at $10^{24} \ \rm{cm}^{-2}$, and $90^{\circ}$,
respectively.  For the simulations of the reflection component 2,
${\theta}_{\rm i}$ was fixed on $\theta_{\rm oa}$+1$^{\circ}$. }
\end{figure}

\begin{figure}[!h]
\includegraphics[angle=270,scale=.40]{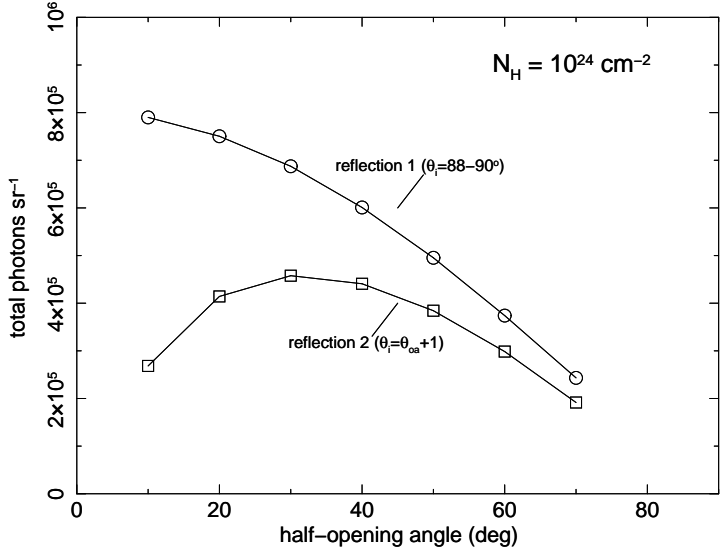}
\caption{Total counts of the reflection components 1 and 2 
as a function of ${\theta}_{\rm oa}$. Open circles and open squares represent the reflection component 1,
 and the reflection component 2, respectively.}
\end{figure}

\begin{figure}[!h]
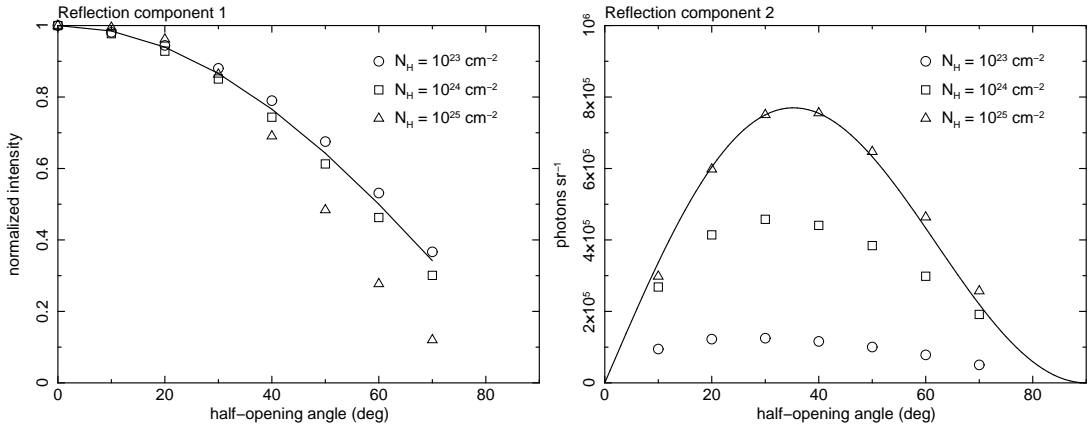

\includegraphics[angle=270,scale=.30]{f8a.eps}
\includegraphics[angle=270,scale=.30]{f8b.eps}
\caption{The $\theta_{oa}$ dependence of the reflection component 1 (left) and 2 (right) for
various column densities, $N_{\rm H}$=10$^{23}$, 10$^{24}$, and 10$^{25}$ cm$^{-2}$.  
The left panel shows the normalized count rate divided by that simulated in the spherical
distributed matter. 
The solid lines in the left and right panels show the covering factor of the torus, $\cos \theta_{\rm oa}$,
and the combination of the covering factor and the apparent size of the visible inner wall,
$\cos \theta_{\rm oa}$ $\cos (2 \theta_{\rm oa}-\pi/2)$. }
\end{figure}

%inclination angle

\begin{figure}[!h]
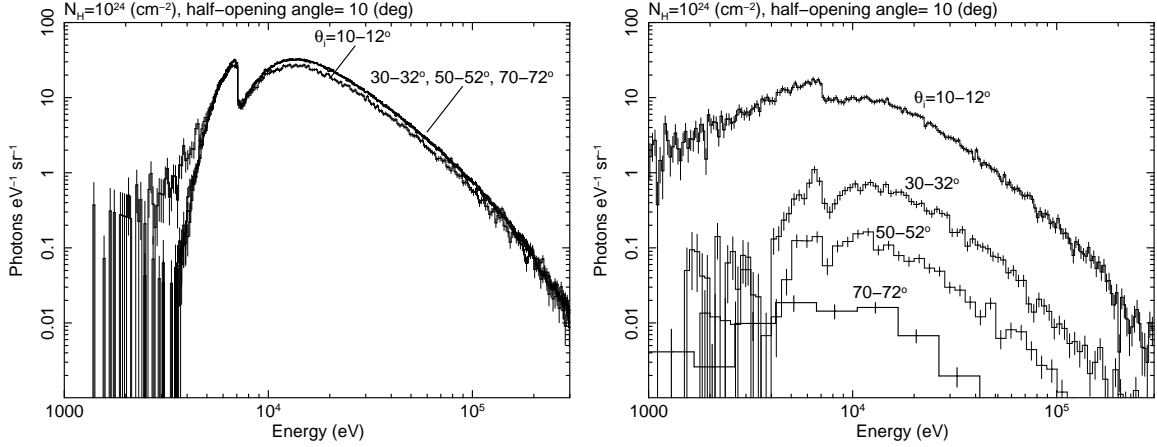

\includegraphics[angle=270,scale=.30]{f9a.eps}
\includegraphics[angle=270,scale=.30]{f9b.eps}
\caption{The ${\theta}_{\rm i}$-dependence of the reflection components 1 (left) and 2 (right). In these simulations, 
$N_{\rm{H}}$ and ${\theta}_{\rm oa}$ were set at $10^{24} \ \rm{cm}^{-2}$ and $10^{\circ}$,
respectively. }
\end{figure}

\begin{figure}[!h]
\includegraphics[angle=270,scale=.40]{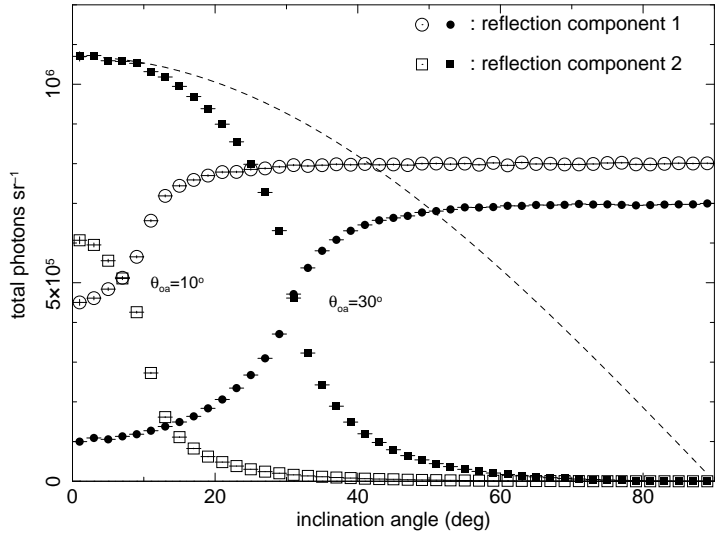}
\caption{Total counts of the reflection components 1 and 2 in the 1~keV--300~keV band as a function of  
${\theta}_{\rm i}$.  Open circles and squares show the total counts of the reflection component 1 and 2 
at ${\theta}_{\rm oa}$=$10^{\circ}$, respectively. We also show the total counts for ${\theta}_{\rm oa}$=
$30^{\circ}$ by closed circles and squares. A curve of cos $\theta_ {\rm i}$ is plotted in the dotted line, 
in order to show a $\theta_ {\rm i}$-dependence of the projected area. }
\end{figure}

%r_in
\begin{figure}[!h]
\includegraphics[angle=270,scale=.40]{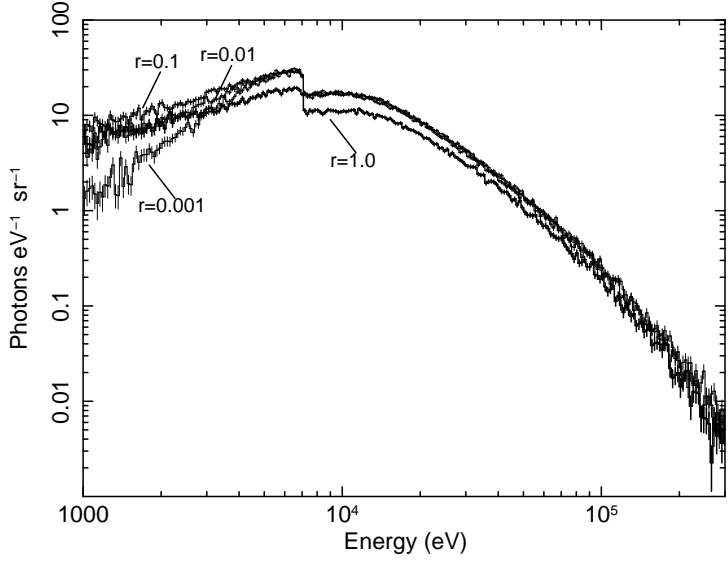}
\caption{Simulated spectra of the reflection component 2 for different
values of the ratio  
$r=r_{\rm{in}}$/$r_{\rm{out}}$. }
\end{figure}

%r_in
\begin{figure}[!h]
\includegraphics[angle=270,scale=.80]{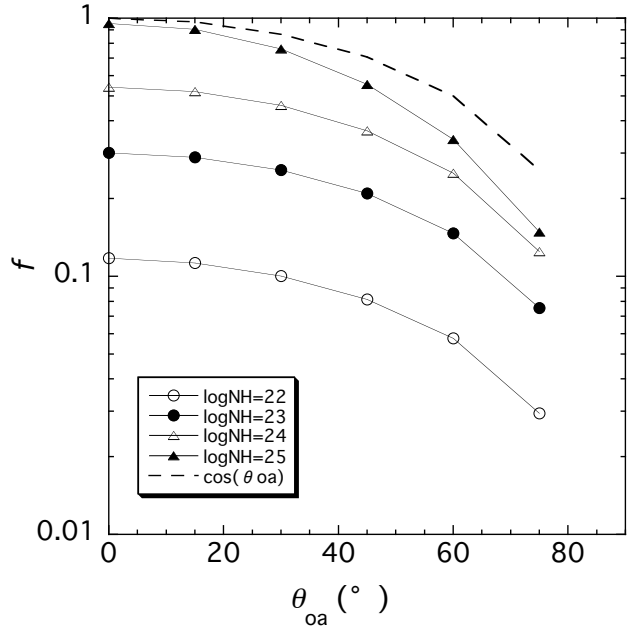}
\caption{The fraction of the absorbed luminosity with respect to the intrinsic source luminosity.
The fraction was estimated for $N_{\rm H}$=10$^{22}$, 10$^{23}$, 10$^{24}$, and 10$^{25}$ cm$^{-2}$.
We also plot a curve of $\cos$ $\theta_{\rm oa}$ in the dashed line, in order to display the covering factor of
the dusty torus. }
\end{figure}

%fe line

\begin{figure}
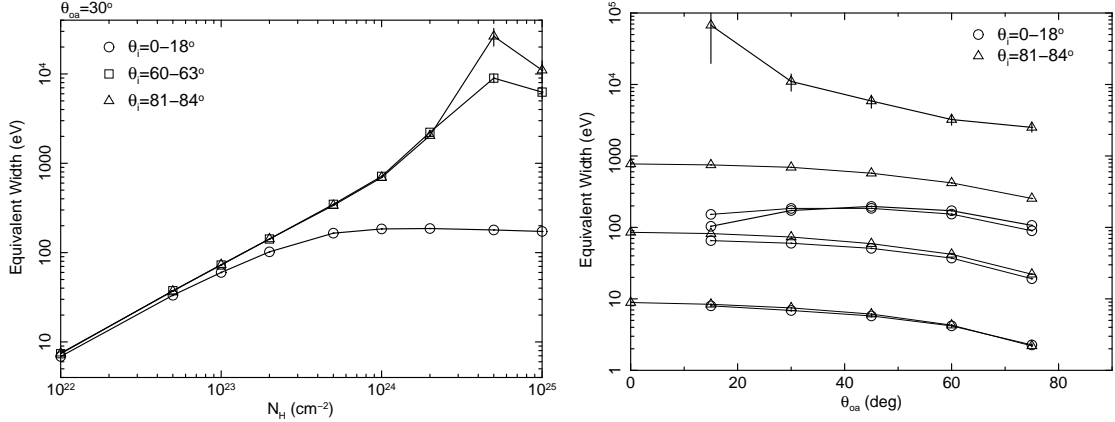

\includegraphics[angle=270,scale=.30]{f13a.eps}
\includegraphics[angle=270,scale=.30]{f13b.eps}
\caption{Equivalent width of the iron K$\alpha$ line to the total continuum emission,
summed over the direct and reflection components. The left panel shows the $N_{\rm H}$ dependence
of  the equivalent width for $\theta_{\rm oa}$=30$^{\circ}$, and the right panel shows
the dependence on $\theta_{\rm oa}$ for $N_{\rm H}$=10$^{22}$, 10$^{23}$, 10$^{24}$, and 10$^{25}$ cm$^{-2}$
from the bottom to the top (or from thin to thick lines).  The open circles, squares, and triangles indicate the equivalent width 
in the ranges $\theta_{\rm i}$=0--18$^{\circ}$, 60--63$^{\circ}$, and 81--84$^{\circ}$, respectively.  
We plot the equivalent width for the spherical distribution at $\theta_{\rm oa}$=0, instead of using
our simple torus model with $\theta_{\rm oa}$=0.}
\end{figure}

\begin{figure}[!h]
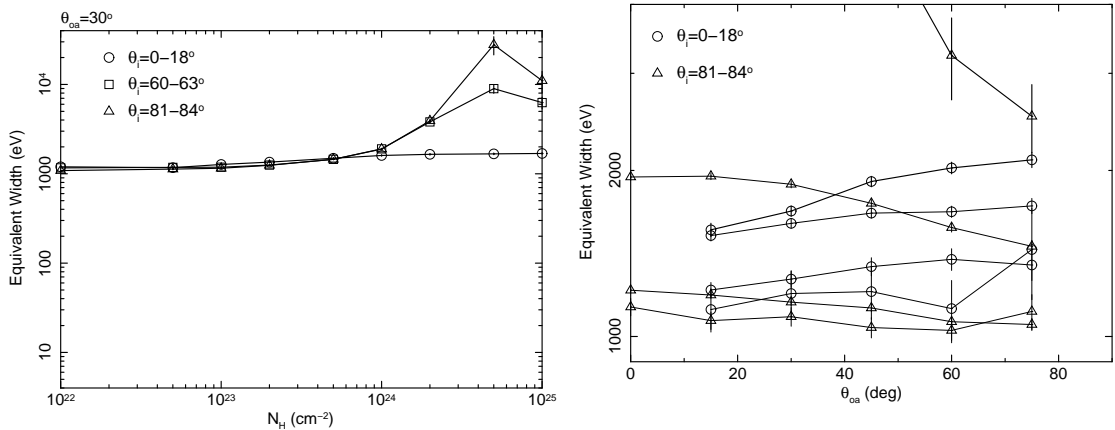

\includegraphics[angle=270,scale=.30]{f14a.eps}
\includegraphics[angle=270,scale=.30]{f14b.eps}
\caption{Equivalent width of the iron K$\alpha$ line to the reflection components 1+2. 
Symbols are the same as in Figure 12.  }
\end{figure}

%ref2_vs_oa
\begin{figure}[!h]
\includegraphics[scale=.40]{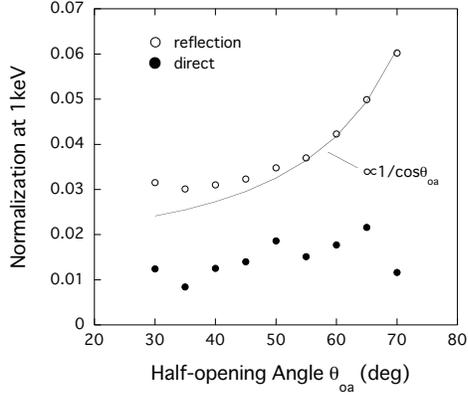}
\caption{The normalization of the direct and reflection components as a function of $\theta_{\rm oa}$.
$\theta_{\rm i}$=$\theta_{\rm oa}$+1$^{\circ}$ is assumed.
The closed and open circles display their normalizations as shown in the figure.
The solid line indicates a curve proportional to 1/$\cos \theta_{\rm oa}$. }
\end{figure}

\begin{figure}[!h]
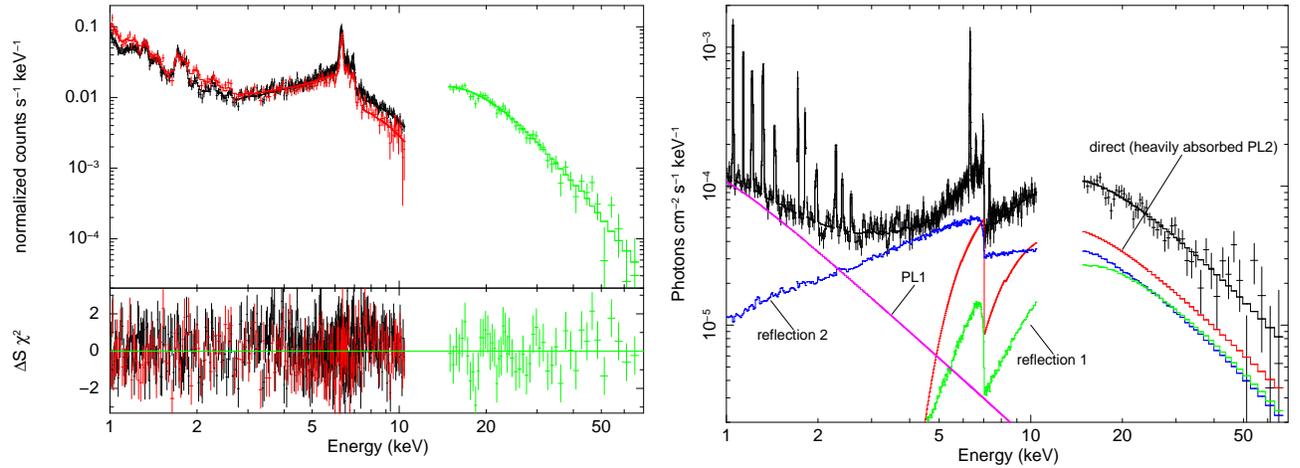

\begin{minipage}[t]{.47\textwidth}
  \includegraphics[angle=270,scale=.35]{f16a.eps}
 \end{minipage}
 \hfill
 \begin{minipage}[t]{.47\textwidth}
  \includegraphics[angle=270,scale=.35]{f16b.eps}

 \end{minipage}
\caption{The wide-band spectrum of Mrk 3 observed with Suzaku fitted with our model (left), and 
the unfolded X-ray spectrum (right).
The spectra in the left panel are obtained by the XIS-FI (black), XIS-BI (red), and HXD-PIN (green). 
The energy range around the Si-K edge (1.875-1.845 keV) is ignored for spectral 
fitting. The right panel shows continuum emission
from PL1, reflection 1, reflection 2 and heavily absorbed PL2.}
\end{figure}

%residual
%\begin{figure}[!h]
%\includegraphics[angle=270,scale=.40]{image_forT/fig17.eps}
%\caption{Residuals of the observed data from the best-fit models. 
%Panels (a), (b), (c), (d), and (e) show the residuals for the best-fit models fixed on $N_{\rm H2}$ = 1, 3, 5, 7, 10$\times$10$^{24}$ cm$^{-2}$, respectively.
%We plotted only XIS-FI and HXD-PIN data}
%\end{figure}

% Tables

\clearpage

\begin{table}
\begin{center}
\caption{Parameters and grids of table model}
\footnotesize
\begin{tabular}{cc}
\tableline\tableline
Parameters & Parameter grids \\
\tableline
Photon index & 1.5, 1.9, 2.5\\
$N_{\rm H}$ ($\times$10$^{22}$ cm$^{-2}$) & 1, 5, 10, 50, 100, 200, 300, 500, 700, 1000\\
half-opening angle ($^{\circ}$) & 0, 10, 20, 30, 40, 50, 60, 70 \\
inclination angle ($^{\circ}$)$^{a}$ &  1-- 89  in steps of 2 \\
\tableline
\end{tabular}
%% Any table notes must follow the \end{tabular} command.
\tablenotetext{a}{We selected X-rays within the inclination angle$\pm$ 1$^{\circ}$.}
\end{center}
\end{table}

%\label{tbl-1}

\begin{table}
\begin{center}
\caption{$\chi^{2}$ on the parameter grid of $\theta_{\rm oa}$ and $\theta_{\rm i}$.}
\begin{tabular}{cccccccccc}
\tableline\tableline
$\theta_{i}-\theta_{\rm oa}$ & \multicolumn{9}{c}{$\theta_{\rm oa}$}\\
                                          & 30$^{\circ}$ & 35$^{\circ}$ & 40$^{\circ}$ & 45$^{\circ}$ & 50$^{\circ}$ & 55$^{\circ}$ & 60$^{\circ}$ & 65$^{\circ}$ & 70$^{\circ}$ \\
\tableline
%%-2 & \it{706.6} & 706.5      & \it{705.4} & 709.3      & \it{700.4} & 700.3      & \it{707.7} & 706.9       & \it{708.1}\\
%%-1 &  709.3     & \it{709.8} & 706.5 & \it{708.4} & 701.2      & \it{705.7} & 709.1       & \it{704.9} & 708.2\\
%% 0 & \it{705.9} & 716.4      & \it{706.2} & 708.6      & \it{701.8} & 712.8      & \it{708.8} & 702.5       & \it{710.4}\\
%%+1 & 710.6     & \it{718.5} & 713.1 & \it{712.7} & 708.8      & \it{713.5} & 711.0      & \it{708.9} & 729.4\\
%%+2 & \it{722.7} & 742.3      & \it{740.1} & 738.7      & \it{730.4} & 739.8      & \it{737.1} & 745.3      & \it{753.6}\\
-2 &  704.6 & 706.7 & 706.2 & 702.1  & 701.9 & 700.8  & 706.6  & 703.3   & 706.3 \\
-1 &  701.6 & 705.9 & 705.7 & 702.8  & 700.2 & 703.2  & 708.9  & 704.0   & 706.2\\
 0 &  701.7 & 705.3 & 706.2 & 704.1  & 702.9 & 707.3  & 707.0  & 706.7   & 707.6\\
+1 & 705.1 & 710.2 & 711.5 & 708.8  & 711.8 & 710.2  & 713.4  & 713.6   & 723.6\\
+2 & 724.9 & 736.4 & 724.4 & 734.1  & 729.3 & 737.6  & 737.7  & 752.0   & 753.6\\
\tableline
\end{tabular}
%% Any table notes must follow the \end{tabular} command.
%\tablenotetext{a}{Sample footnote for table~\ref{tbl-2} that was
%generated with the \LaTeX\ table environment}
\end{center}
\end{table}
%\label{tbl-2}

\begin{table}
\begin{center}
\caption{Best-fit model parameters.}
\footnotesize
\begin{tabular}{cccccccc}
\tableline\tableline
Photon Index & $N_{\rm H1}$ & $L_{\rm 2-10}^{\rm direct}$ & $N_{\rm H2}$ & $L_{\rm 2-10}^{\rm refl}$  & ${\theta}_{\rm oa}$ & ${\theta}_{\rm i}$  & ${\chi}^{2}$/(d.o.f.) \\
   &  (${\times}$$10^{24}$ $\rm{cm^{-2}}$)  & ( erg s$^{-1}$ ) &  (${\times}$$10^{24}$ $\rm{cm^{-2}}$) & ( erg s$^{-1}$ ) & ($^{\circ}$) &  ($^{\circ}$)  &  \\
\tableline
1.82 & 1.1& 2.1$\times10^{43}$  &  1.5 &  5.1$\times10^{43}$   &50(fixed) & 51(fixed) &726/613\\
\tableline
%\tablecomments{$N_{\rm H1}$ was fixed at the Galactic column density.}
\multicolumn{6}{l}{Note. --- $N_{\rm H1}$ was linked to $N_{\rm H2}$.}
\end{tabular}
%% Any table notes must follow the \end{tabular} command.
%\tablenotetext{a}{Sample footnote for table~\ref{tbl-2} that was
%generated with the \LaTeX\ table environment}
\end{center}
\end{table}

%\label{tbl-3}

\pagebreak
\clearpage




\clearpage

\end{document}